\documentclass[12pt]{article}
\usepackage{amsmath}
\usepackage{graphicx,psfrag,epsf}
\usepackage{enumerate}
\usepackage{url}
\usepackage{graphicx}
\usepackage{hyperref}
\usepackage{float}
\usepackage[square,numbers]{natbib}
\newcommand{\blind}{0}

\def\spacingset#1{\renewcommand{\baselinestretch}%
{#1}\small\normalsize} \spacingset{1}

  \title{\bf Assessing the XDC Network: A Comprehensive Evaluation of its qualitative and technical aspects}
  \author{
  Atul Khekade\\
  \texttt{atul@xinfin.org}
  \and
  Omkar Mestry\\
  \texttt{Omkar@xinfin.org}
  \and
  Van Khanh Nguyen\\
  \texttt{13520392@ms.uit.edu.vn}
}
\begin{document}

  \maketitle

\if1\blind
{

  \begin{center}
\end{center}
  \medskip
} \fi

\bigskip
\begin{abstract}
This research provides a thorough assessment of the XDC Network, a delegated proof of stake (XDPoS) consensus-based blockchain technology, across its technical, security, and business dimensions. The study evaluates the network’s decentralization, scalability, and security features, including its Nakamoto coefficient, validator participation, and client distribution. Additionally, it examines the developer ecosystem, including GitHub metrics, and business aspects such as transaction costs and predictability. The findings of this research will provide valuable insights into the strengths and weaknesses of the XDC Network, informing stakeholders and decision-makers about its suitability for various use cases, particularly in trade finance, asset tokenization, and enterprise blockchain solutions.
\end{abstract}

\newpage
\spacingset{1.45}
\section{Introduction}
\label{sec:intro}

The XDC Network, a decentralized, open-source blockchain technology, has gained significant attention in recent years for its potential to transform various industries, including trade finance, asset tokenization, and enterprise blockchain solutions. As the blockchain landscape continues to evolve, it is essential to evaluate the XDC Network’s technical, security, and business aspects to understand its capabilities, limitations, and potential areas for improvement.

This research aims to provide a comprehensive assessment of the XDC Network using a structured framework that evaluates its Distributed Ledger Technology (DLT) across four three categories: Network, Developer Ecosystem, and Business. This framework will provide a detailed examination of the XDC Network’s decentralized architecture, security features, scalability, smart contract platform, interoperability, masternode distribution, network participation, client distribution, developer ecosystem, cost structure, and potential use cases.

By applying this framework to the XDC Network, this research will provide a thorough understanding of its strengths, weaknesses, opportunities, and threats, ultimately informing stakeholders, developers, and users about its suitability for various use cases and its potential for future growth and adoption.

\section{Source Data and Processing Methods}
\label{sec:data}
This research collects data from the following sources:
\begin{itemize}
    \item XDC Nodes information\cite{XDCnodes}
    \item XDC Ecosystem Developer Report\cite{XDCDeveloperReport}
    \item XDC Github Repositories\cite{XDCGithub}
    \item XDCScan.io\cite{XDCScan}
\end{itemize}
The XDC Nodes information has been processed and enriched with data from \href{infobyip.com}{infobyip.com} to include the geographic locations of masternodes, resulting in a "Node Data Enriched" dataset\cite{processeddata}.

Additionally, a Python script was used to extract GitHub statistics, yielding the "XDC Repo Stats" dataset\cite{processeddata} and another script is also used to parse \cite{XDCFees} to create the charts.

\newpage
\section{Network Assessment}
\label{sec:network}

Network decentralization
    \begin{itemize}
        \item \textbf{Nakamoto Coefficient}: To calculate the Nakamoto Coefficient\cite{nakamotocoefficient} for the XDC Network, we consider its XDPoS 2.0 consensus algorithm\cite{XDCNetworkDocumentation}. In this system, 108 masternodes are randomly selected for each epoch, and the HotStuff protocol is then applied, which requires a super-majority of 2/3 of the elected masternodes to finalize a block. Therefore, the Nakamoto Coefficient of the XDC Network is calculated as:
\[
\text{Nakamoto Coefficient} = \left\lfloor \frac{2}{3} \times 108 \right\rfloor = 72
\]
A Nakamoto Coefficient of 72 in a network of 429 nodes indicates a high degree of decentralization, enhancing the network’s resistance to centralized control and malicious attacks. This structure ensures no single entity can dominate the network, providing a secure and trustworthy environment for users. Decentralization also promotes a democratic decision-making process, where nodes have a say in the direction of the network.

However, decentralization can also lead to slower decision-making and coordination, hindering the network’s ability to adapt to changing circumstances. Additionally, decentralization can lead to a lack of accountability. Despite these drawbacks, the benefits of decentralization in this context appear to outweigh the drawbacks, providing a robust defense against centralized control and malicious attacks.
    \end{itemize}
Participation Rate and Number of Validators:
    \begin{itemize}
        \item \textbf{Participation Rate:} The participation rate can be calculated by determining the ratio of Masternodes with active synchronization status to those with inactive synchronization status. 
        \begin{table}[h]
\centering
\begin{tabular}{|c|c|}
\hline
\textbf{Syncing} & \textbf{Masternodes} \\
\hline
FALSE & 19 \\
TRUE & 410 \\
\hline
\textbf{Grand Total} & 429 \\
\hline
\end{tabular}
\caption{Syncing status of Masternodes}
\end{table}

    Based on the table, we have the following participation rate calculation::
\[
    \text{Participation rate} = \left(\frac{410}{429}\right) \times 100 \approx 95.57\%
    \]

        \item \textbf{Number of Validators}: According to XDPoS 2.0 consensus algorithm\cite{XDCNetworkDocumentation}, there are 108 masternodes acting as validators.
    \end{itemize}
Client Distribution
    \begin{itemize}
        \item \textbf{Geo Distribution}: The table below displays the geographical distribution, showcasing data enriched from XDC nodes\cite{XDCnodes}, further enhanced with information from InfoByIp\cite{infobyip}.
\begin{table}[H]
    \centering
    \begin{tabular}{|l|l|}
    \hline
        \textbf{Country} & \textbf{Masternodes} \\ \hline
        Australia & 10 \\ \hline
        Belgium & 1 \\ \hline
        Canada & 8 \\ \hline
        China & 1 \\ \hline
        Finland & 7 \\ \hline
        France & 8 \\ \hline
        Germany & 132 \\ \hline
        Hong Kong & 1 \\ \hline
        Japan & 7 \\ \hline
        Poland & 2 \\ \hline
        Singapore & 129 \\ \hline
        Slovenia & 1 \\ \hline
        Taiwan & 1 \\ \hline
        The Netherlands & 16 \\ \hline
        United Kingdom & 27 \\ \hline
        United States & 78 \\ \hline
        \textbf{Grand Total} & 429 \\ \hline
    \end{tabular}
    \caption{Geo distribution of masternodes}

\end{table}
    \end{itemize}
\bigskip

The masternode distribution showcases a widespread presence across various regions, highlighting the benefits of geo-distribution in promoting network resilience, lowering latency, and navigating regulatory frameworks. Notably, Germany and Singapore have a high concentration of nodes, indicating a significant reliance on these economically stable and technologically advanced jurisdictions.

However, this concentration also poses potential risks if these countries were to experience sudden regulatory changes or localized disruptions. Despite this, the network appears to be well-distributed, but there is still room for improvement in terms of geographical spread to fully leverage the advantages of a decentralized network structure.
\newline
Overall, while the network appears to be well-distributed, there is still room for improvement in terms of geographical spread to fully leverage the advantages of a decentralized network structure.
\newpage
    \begin{itemize}
        \item \textbf{Host Distribution}: The table below displays the host distribution, showcasing data enriched from XDC nodes\cite{XDCnodes}, further enhanced with information from InfoByIp\cite{infobyip}.
\begin{table}[H]
    \centering
    \begin{tabular}{|l|l|}
    \hline
        Host & Masternodes \\ \hline
        Singapore-based hosts & 129 \\ \hline
        Germany-based hosts & 132 \\ \hline
        US-based hosts & 78 \\ \hline
        United Kingdom & 27 \\ \hline
        Others & 63 \\ \hline
        \textbf{Grand Total} & 429 \\ \hline
    \end{tabular}
    \caption{Host distribution of masternodes}
\end{table}

The masternode hosting data reveals a diverse range of platforms, mitigating platform-specific risks and centralization. With 429 nodes spread across various services, the network is resilient against outages or disruptions. This diversity allows for cost optimization and enhances network robustness.

However, the concentration of nodes in a few locations, such as Singapore and Germany, raises concerns about centralization risks and influence on network operations. To further optimize the network's resilience and efficiency, it's essential to reduce dependencies on these providers and increase presence in underutilized services.
\end{itemize}
\begin{itemize}
    \item \textbf{Client Diversity}:  The following tables display the host distribution, showcasing data enriched from XDC nodes\cite{XDCnodes}.
\begin{table}[!ht]
    \centering
    \begin{tabular}{|l|l|}
    \hline
        client version & Masternodes \\ \hline
        0.1.1 & 429 \\ \hline
        \textbf{Grand Total} & 429 \\ \hline
    \end{tabular}
    \caption{Client versions of masternodes}
\end{table}
\begin{table}[!ht]
    \centering
    \begin{tabular}{|l|l|}
    \hline
        node version & Masternodes \\ \hline
        XDC/v1.4.4-stable/linux-amd64/go1.14.15 & 3 \\ \hline
        XDC/v1.4.6-stable/linux-amd64/go1.14.15 & 1 \\ \hline
        XDC/v1.4.6-stable/linux-amd64/go1.17.13 & 1 \\ \hline
        XDC/v1.5.0/linux-amd64/go1.21.5 & 1 \\ \hline
        XDC/v1.6.0/linux-amd64/go1.21.10 & 414 \\ \hline
        XDC/v1.6.0/linux-amd64/go1.21.11 & 1 \\ \hline
        XDC/v1.6.0/linux-amd64/go1.22.0 & 1 \\ \hline
        XDC/v1.6.0/linux-amd64/go1.22.4 & 1 \\ \hline
        XDC/v1.6.0/windows-amd64/go1.21.7 & 4 \\ \hline
        XDC/v2.2.0-beta1/linux-amd64/go1.21.11 & 1 \\ \hline
        XDC/v2.3.0-beta1-de2be8fa/linux-amd64/go1.21.12 & 1 \\ \hline
        \textbf{Grand Total} & 429 \\ \hline
    \end{tabular}
    \caption{Node versions of masternodes}
\end{table}
\begin{table}[!ht]
    \centering
    \begin{tabular}{|l|l|}
    \hline
        os & Masternodes \\ \hline
        linux & 425 \\ \hline
        windows & 4 \\ \hline
        \textbf{Grand Total} & 429 \\ \hline
    \end{tabular}
    \caption{OS versions of masternodes}
\end{table}
\newline
\newline
On one hand, the uniformity of client and node versions can be beneficial in terms of simplifying maintenance and ensuring consistency across the network. With all nodes running the same version, it’s easier to roll out updates and patches, and developers can focus on a single codebase. Additionally, the dominance of Linux as the OS of choice can be seen as a positive, as it’s a widely-used and well-supported platform.

On the other hand, this lack of diversity in client and node versions increases the risk of network-wide exploits or failures if a vulnerability is found. If a single vulnerability is discovered, it could potentially affect the entire network, given the uniformity of the versions. Similarly, the high degree of homogeneity in OS usage could limit the network’s resilience against OS-specific vulnerabilities. To mitigate these risks, it’s essential to encourage diverse client and node versions, as well as support multiple operating systems, to promote a more robust and resilient infrastructure.
\end{itemize}
\section{Developer Ecosystem Assessment}
\label{sec:developer}
\begin{itemize}
    \item \textbf{GitHub Statistics}: As of July 22, 2024, the statistics for all repositories belonging to the XDC Network are shown in the table below.
\begin{table}[!ht]
    \centering
    \begin{tabular}{|l|l|}
    \hline
        Category & Total/Size \\ \hline
        Stars & 350 \\ \hline
        Open Issues & 320 \\ \hline
        Forks & 678 \\ \hline
        Repository Size & 4164429 \\ \hline
        Watchers & 350 \\ \hline
        Subscribers & 222 \\ \hline
        Network Count & 98382 \\ \hline
    \end{tabular}
\end{table}

The GitHub statistics suggest a promising outlook for the project, with 350 stars and watchers indicating moderate popularity and ongoing interest. The 678 forks and a network count of 98,382 demonstrate significant collaborative interest and influence, highlighting the project’s potential for growth and development. Additionally, the substantial repository size of 4,164,429 suggests a mature, complex codebase that has been thoroughly developed and refined.

However, the 320 open issues may indicate a need for enhancements or bug fixes, which could affect the project’s perceived stability and responsiveness. Furthermore, the relatively low number of 222 subscribers compared to stars and forks suggests a smaller core group committed to ongoing engagement, which may impact the project’s ability to sustain momentum and drive progress. Despite these areas for improvement, the overall GitHub statistics paint a positive picture of the project’s standing within the developer community.

\end{itemize}

\section{Business Assessment}
\label{sec:bussiness}
\begin{itemize}
    \item \textbf{Cost of Transaction and Predictability}
\end{itemize}
\includegraphics[width=\linewidth]{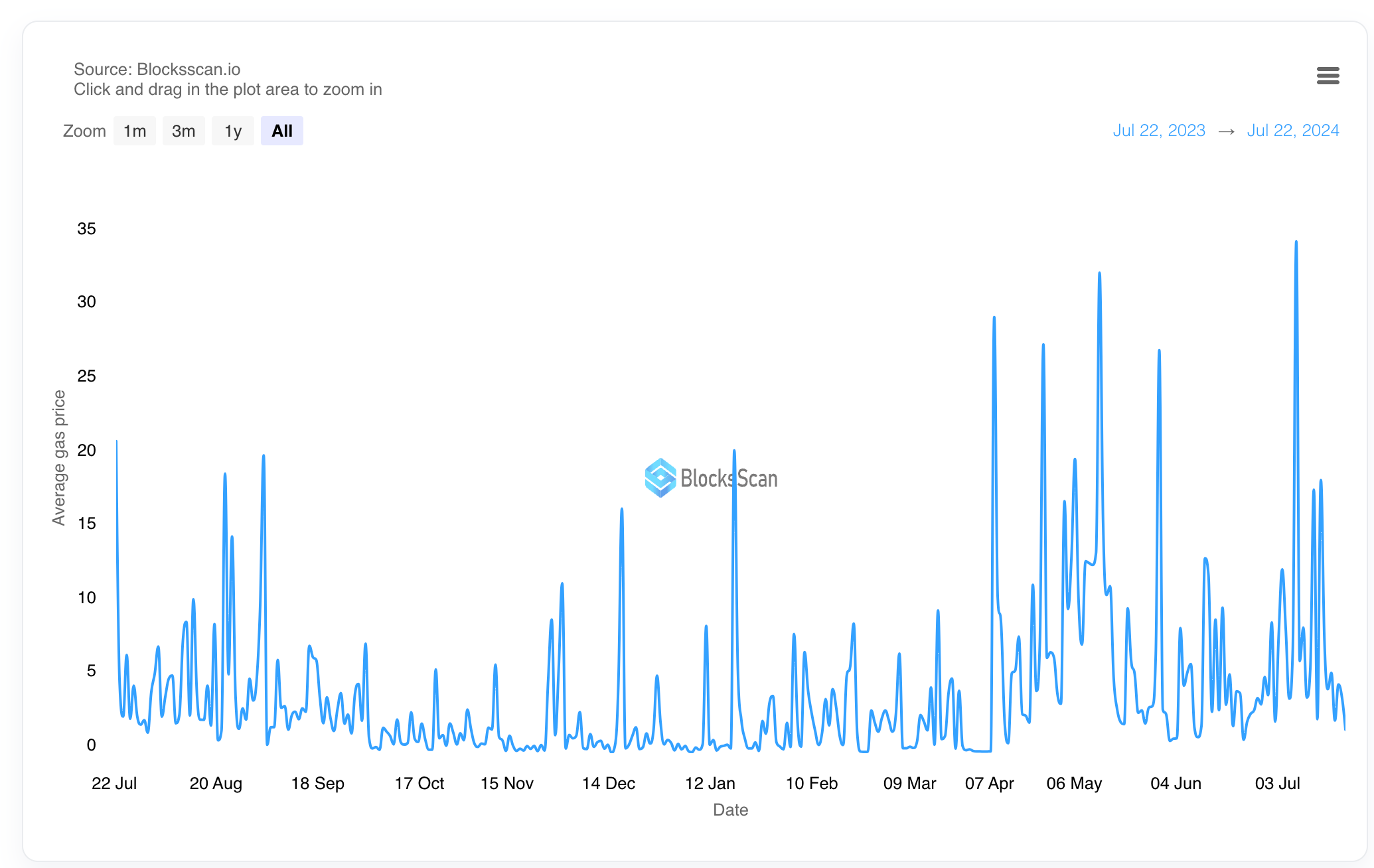}
The Gas price graph from XDCScan reveals significant variability in gas prices, with sharp peaks exceeding 25 Gwei, indicating high network demand or reduced capacity. This leads to increased transaction costs, deterring frequent or small transactions. The frequent fluctuations in the graph make it challenging to predict transaction costs, negatively affecting user experience.

Meanwhile, the network can take steps to mitigate these issues and improve user experience. To attract and retain users, the network needs to balance low and predictable gas prices. Implementing mechanisms like dynamic block sizing or efficient transaction processing can help stabilize fees and improve user experience. By doing so, the network can reduce the negative impact of high and unpredictable costs, which may drive users to alternative networks.
\begin{itemize}
    \item \textbf{Transaction fees (in XDC):}
\end{itemize}
\includegraphics[width=\linewidth]{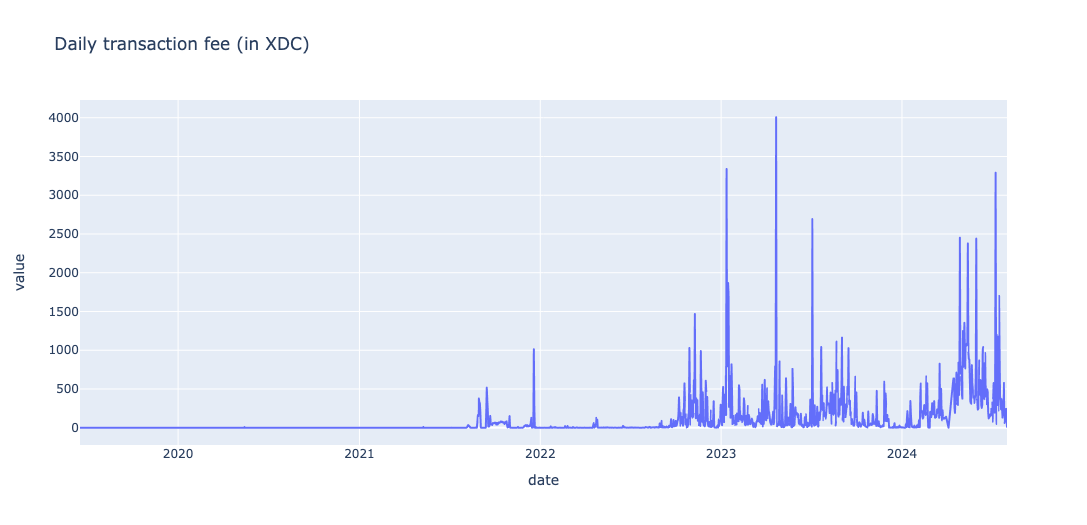}
The line graph illustrates the daily transaction fees in XDC from early 2020 to mid-2024, revealing a trend of increasing volatility over time. Initially, from 2020 to early 2021, transaction fees were consistently low, suggesting minimal network activity or stable fee policies. This period of stability may have been beneficial for users, as it provided a predictable and affordable environment for transactions.

However, the graph also shows a noticeable increase in variability and peak fees emerging in 2021 and continuing through 2022, with fees occasionally reaching around 1000 XDC. The volatility intensifies in 2023, with more frequent and higher spikes, indicating possible network congestion or changes in fee structures. By 2024, the trend of sharp spikes persists, with fees peaking between 3000 and 4000 XDC, reflecting ongoing adjustments or increased demand within the network. This increased volatility may have negatively impacted user experience, making it more challenging to predict and budget for transaction fees.

\includegraphics[width=\linewidth]{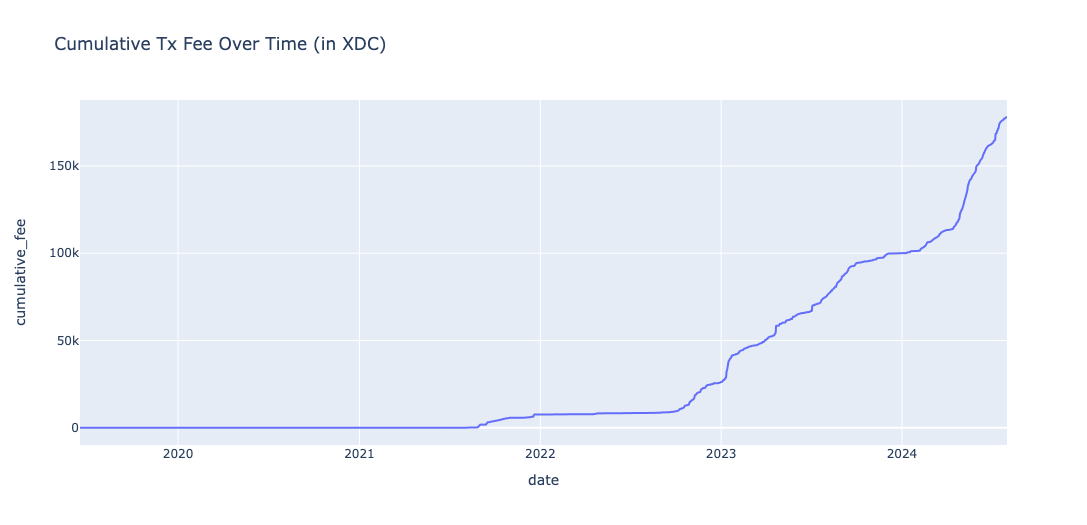}
The line graph displays the cumulative transaction fees in XDC from the start of 2020 to mid-2024, showing minimal growth up to 2021, indicating a low level of transaction activity. This initial period of slow growth may have been beneficial for the network, as it allowed for steady development and refinement without excessive strain on resources.

In contrast, from 2022, a steady increase begins, suggesting a gradual rise in network usage or transaction costs. The growth in cumulative fees becomes more pronounced from 2023 onwards, accelerating significantly and reaching 177857.76288214893 XDC by the end of July 2024. This sharp increase likely reflects heightened transaction activity, possible network developments, or changes in fee structures during this later period, which may have had both positive and negative impacts on the network and its users.

\bigskip

\bibliography{references.bib}

\begin{thebibliography}{1}

\bibitem{XDCDeveloperReport}
Electric Capital.
\newblock Xdc ecosystem developer report.
\newblock \url{https://www.developerreport.com/ecosystems/xdc}, 2024.
\newblock Accessed: 2024-07-20.

\bibitem{infobyip}
infobyip.com.
\newblock Domain and ip bulk lookup tool.
\newblock \url{https://www.infobyip.com/}, 2024.

\bibitem{XDCGithub}
XDC Network.
\newblock Xdc github repositories.
\newblock \url{https://github.com/XinFinOrg}, 2024.
\newblock Accessed: 2024-07-20.

\bibitem{XDCNetworkDocumentation}
XDC Network.
\newblock Xdc network documentation.
\newblock \url{https://docs.xdc.community/}, 2024.
\newblock Accessed: 2024-07-20.

\bibitem{XDCnodes}
XDC Network.
\newblock Xdc nodes data.
\newblock \url{https://stats-ws.xdc.org/get-table-nodes}, 2024.
\newblock Accessed: 2024-07-20.

\bibitem{nakamotocoefficient}
Balaji~S. Srinivasan and Leland Lee.
\newblock Quantifying decentralization.
\newblock \url{https://news.earn.com/quantifying-decentralization-e39db233c28e}, 2017.

\bibitem{processeddata}
Up.
\newblock Processed dataset.
\newblock \url{https://drive.google.com/drive/folders/1_AN5rbhyopXWvhzIDoF46Wr2y4-1cLgt?usp=drive_link}, 2024.

\bibitem{XDCScan}
XDCScan.
\newblock Xdcscan.io - xdc gas prices.
\newblock \url{https://xdcscan.io/chart/txnsFee}, 2024.
\newblock Accessed: 2024-07-20.

\bibitem{XDCFees}
XDCScan.
\newblock Xdcscan.io - xdc transaction fees.
\newblock \url{https://api.xdcscan.io/lines/txnsFee?from=2017-01-01&to=2024-08-01}, 2024.
\newblock Accessed: 2024-07-31.

\end{thebibliography}

\end{document}